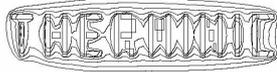

# Acoustically Enhanced Boiling Heat Transfer

Z. W. Douglas, M. K. Smith, A. Glezer
Woodruff School of Mechanical Engineering
Georgia Institute of Technology
771 Ferst Drive NW
Atlanta, GA 30332

*Abstract*-An acoustic field is used to increase the critical heat flux (CHF) of a flat-boiling-heat-transfer surface. The increase is a result of the acoustic effects on the vapor bubbles. Experiments are performed to explore the effects of an acoustic field on vapor bubbles in the vicinity of a rigid-heated wall. Work includes the construction of a novel heater used to produce a single vapor bubble of a prescribed size and at a prescribed location on a flat-boiling surface for better study of an individual vapor bubble's reaction to the acoustic field. Work also includes application of the results from the single-bubble heater to a calibrated-copper heater used for quantifying the improvements in CHF.

## I. OVERVIEW

A major challenge in the design and packaging of high-density microelectronic devices and systems (e.g., high-speed microprocessors, optoelectronics, micro- and millimeter-wave power electronics, power converters, etc.) is the ever-increasing demand for high-power-density heat dissipation to ensure efficient and proper operation and to prevent thermal damage. Cooling fluxes as high as 1000 W/cm$^2$ are projected in the next five to ten years for a number of high-power electronic applications (e.g., power conditioning transistors for electronic motor control in hybrid vehicles, [1]). Cooling demands at these power densities present system planners and designers with monumental thermal management problems that will necessitate new approaches and technologies for enhanced heat transfer. Though much progress is being made in areas like single-phase microchannel coolers [2], these demands go well beyond the capabilities of conventional forced air or liquid convection cooling and clearly point towards the use of two-phase cooling technologies [3, 4].

Two-phase thermal management based on submerged boiling heat transfer has received considerable attention in recent years because of its potential to enable high heat flux with relatively simple hardware implementations (e.g., [5]) and because it is relatively easy to couple to the system-level heat transfer. However, this attractive heat transfer approach is hampered by a major drawback that has limited its utilization in practical systems, namely, the CHF limit on the maximum heat transfer. At the CHF, nucleate boiling transitions to film boiling resulting in a vapor layer that insulates the surface and causes a drastic increase in the surface temperature that can severely damage the cooled hardware. The severity of this limitation can be potentially mitigated by cross flow above the heated surface, but at a substantial increase in complexity and cost.

The primary objective of the present work is to investigate a new approach for enhancing boiling heat transfer by using forced detachment of the vapor bubbles from the heated surface and by inhibiting the instabilities that cause the transition from nucleate boiling to film boiling in order to achieve a higher heat flux than in the "natural" limit. The forced vapor-bubble detachment technique builds on earlier experiments at Georgia Tech [6-8] and exploits interfacial excitation of the bubbles to manipulate their contact-line and interfacial dynamics, thereby controlling the detachment of the bubbles from the heated surface.

## II. ACOUSTIC FORCING OF BUBBLES

The effect of an acoustic field on a trapped gas bubble is investigated using an acoustic beam induced by a circular ($D = 31.2$ mm), submerged acoustic driver operating in the kHz range. The acoustic driver is submerged parallel to a horizontal surface at a separation distance of 1 cm. An air bubble is injected beneath the surface (Fig. 1) and held attached to the surface by buoyancy and surface tension. A sufficiently high level acoustic field ($P_a > 0.5$ kPa, where $P_a$ is the sound pressure amplitude) induces capillary waves on the surface of the bubble. The frequency band in which the bubble responds

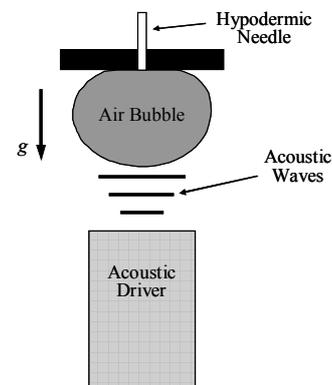

Fig. 1 Schematic of an air bubble injected underneath a horizontal surface in the acoustic field (not shown to scale).

This work was supported by the NASA Microgravity Research Program under Grant NAG3-2763 and the Woodruff Foundation.

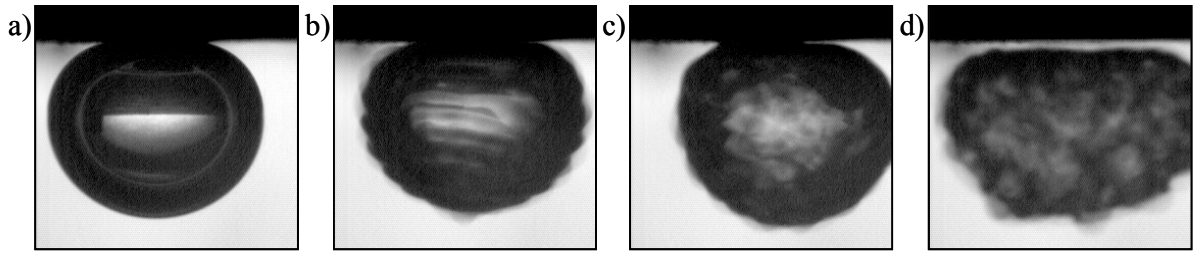

Fig. 2 Series of images from a high speed video showing detachment of an air bubble in a 2 kHz acoustic field. $t/T = 0$ (a), 0.4 (b), 1.1 (c), and 3.7 (d).

depends on the characteristic size of the bubble, and this receptive frequency band is found to broaden as the sound pressure increases. For a given frequency to which the bubble is receptive, as the sound pressure increases ($P_a > 40$ kPa), the capillary waves increase in magnitude until the bubble's surface appears to be completely covered by turbulent waves. The induced surface waves on the gas bubble can become violent enough to break the contact line and detach the bubble from the surface. Fig. 2 is a series of frames from a high speed video showing a 3.2 mm air bubble detaching from a horizontal surface in the presence of an acoustic field. After detachment, the bubble is held close to the surface by buoyancy, but hovers on a thin (0.1 mm) layer of liquid. Because the bubble is no longer pinned by surface tension effects, a relatively small horizontal force is sufficient to move the bubble. It is noteworthy that the bubble becomes fully detached at $t/T = 3.7$, where $t$ is the time since the acoustic driver was turned on and $T$ is a characteristic time that corresponds to the time a capillary waves takes to travel around the circumference of the bubble (the wave speed is a function of the actuation frequency, the liquid density and surface tension). In Fig. 2, $T = 10$ ms.

The effect on the contact line of the bubble is captured using high-speed (5000 fps) video. Fig. 3 shows a quarter section (allowing for better resolution of the contact line) of a 3.5 mm bubble. The initial acoustic waves ($0 < t / T < 0.7$) appear to press the bubble against the surface as seen in Fig. 3b and c. During this initial time, the surface of the bubble is smooth and there is no evidence of capillary waves. The bubble is pressed very close to the surface, though it is not clear whether the radius of the contact line is changed. In Fig. 3d, capillary waves begin to appear on the bubble's surface and propagate towards the contact line. The motion of the interface appears to push a thin layer of liquid against the contact line, slowly separating it from the surface. The separation of the contact line is first visible as a faint gray line in the upper-left of Fig. 3d. From this point the contact line is quickly contracted and the bubble is "peeled" off the surface. In Fig. 3g, the bubble is completely separated from the surface by a 40 μm layer of liquid. The thickness of the liquid layer increases to approximately 100 μm as shown in Fig. 3h. At this point, the separation from the wall is sufficient to allow surface waves to travel along the wall-side surface of the bubble. One of these waves can be clearly seen in Fig. 3h. The bubble hovers on this thin liquid layer until either a horizontal force pushes it off the side of the surface or the acoustic driver is turned off.

The effects of this new type of acoustic forcing on vapor bubbles are investigated using a novel boiling heat transfer surface (Fig. 4) that is designed to produce a *single* vapor bubble of a prescribed size at a predetermined location. The surface consists of a thermally conductive copper pin ($D = 2.4$ mm) surrounded by an insulating stainless steel annulus ($D = 76.2$ mm). The conductive pin provides a thermal via to the surface, resulting in a hot spot on the surface. The power to the heater is controlled such that the temperature of the hot spot is sufficient to nucleate a vapor bubble, while the rest of the surface remains cooler. A dot of hydrophobic coating at the center of the hot spot provides a prescribed nucleation site such that the contact line of the resulting vapor bubble is determined by the boundary of the dot. Experiments with this surface show that the acoustic field, produced by the acoustic driver described above, is quite effective in removing vapor bubbles from an upward-facing horizontal surface. Fig. 5 is a series of frames showing a vapor bubble being removed from the surface of the single bubble heater (the characteristic time is $T = 5$ ms). Fig. 5a shows the 1.5 mm diameter bubble before the acoustic driver is turned on. Figs. 5b-d show the deformation and detachment of the bubble from the surface as a result of the acoustic field. In Fig. 5e the remains of the vapor bubble have completely moved out of the frame, leaving only a few smaller bubbles that were ejected from the surface of the original bubble as a result of the excitation.

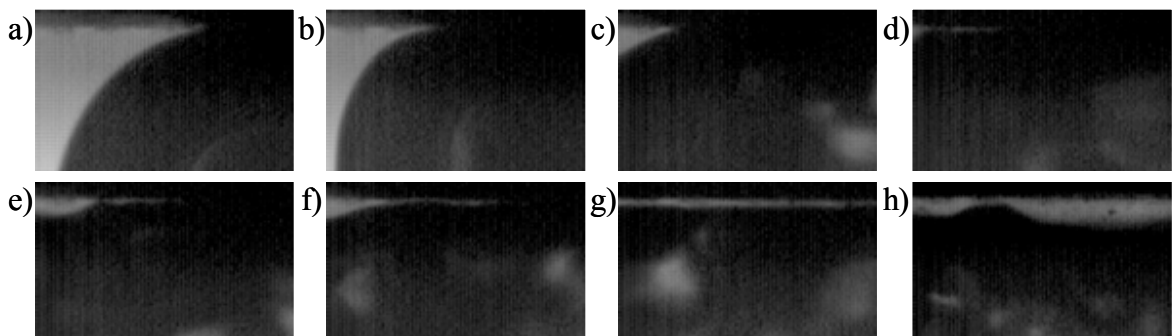

Fig. 3. Frames from a 5000 fps video showing breaking of the contact line and detachment of a bubble. $D = 3.5$ mm, $f = 1.3$ kHz, $P_a = 48.1$ kPa. $t / T = 0$ (a), 0.4 (b), 0.7 (c), 0.8 (d), 0.9 (e), 1.0 (f), 1.2 (g), 2.8 (h).

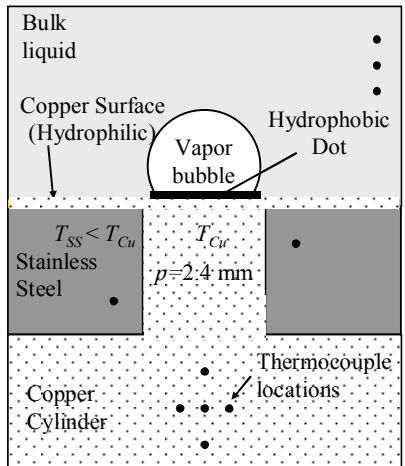

Fig. 4 Close-up schematic of copper pin, stainless steel annulus, electroplated surface and hydrophobic dot (not to scale).

As described earlier, capillary waves are only induced on the surface of a given bubble within a certain band of frequencies. As the sound pressure amplitude increases, this band broadens and the amplitude of the capillary waves increases. The amplitude is at its maximum when the frequency is tuned to the center of this frequency band. One of the goals of the present investigation is to determine the optimal frequency for the removal of a bubble of a given diameter. It is expected that the optimal frequency depends on the diameter, D, the liquid density, ρ, and the surface tension, σ. Dimensional analysis of these parameters produces the dimensionless group α shown in (1). Experimental observations of the parameters that result in bubble removal indicate that α is approximately constant and equal to 6.7. The optimal frequency for bubble removal is plotted in Fig. 6 and shows good agreement with experimental measurements.

$$\left( \frac{D^3 f^2 \rho_\ell}{\sigma} \right) = \alpha \approx 6.7 \tag{1}$$

### III. ENHANCED BOILING

The effectiveness of the induced acoustic field in increasing the CHF of a boiling surface and delaying the transition to film boiling (and the ensuing increase in surface temperature) is investigated using measurements of surface temperature and heat flux through a 1 cm² calibrated copper heater. The exposed surface of the heater is placed at the bottom of a water tank and the acoustic driver is suspended in the liquid, 1 cm above and parallel to the surface of the heater. Preliminary measurements have already demonstrated a significant increase in the CHF. As shown in Fig. 7, the maximum heat flux increases from 333 W/cm² to 404 W/cm², (about 20%). The subcooling is defined as the difference between the bulk temperature and saturation temperature of the liquid. For the two sets of measurements shown, the pressure was atmospheric and the bulk water temperatures were 80.2°C and 86.8°C in the absence and presence of acoustic actuation, respectively. Since the CHF is known to be directly related to the subcooling [9], the measured enhanced CHF is lower than it would have been had the subcooling been held constant and the improvement due to the acoustic actuation is understated. To accurately state the acoustic enhancement it is necessary to determine the equivalent enhanced CHF at a subcooling of 19.8°C. When a correction [9] is applied to account for the difference in the amount of subcooling between the two curves, the improvement in CHF is 34%.

The acoustic field appears to delay the CHF by impeding the formation of large vapor columns and inhibiting the instability that results in the collapse of these vapor columns into a film. Fig. 8a is a photograph of the heated surface undergoing natural boiling (i.e., in the absence of acoustic actuation) immediately prior to the maximum heat flux on the boiling curve, the surface superheat is 44°C in Fig. 7 (the edge of the piezoelectric driver is visible at the top of the image). Fig. 8b shows this same surface at the same temperature, but in the presence of acoustic actuation. These images clearly show the differences in the structures of vapor clusters between the two cases. In Fig. 8b surface waves are visible on the leftmost vapor bubble and the large vapor column seen in Fig. 8a is absent.

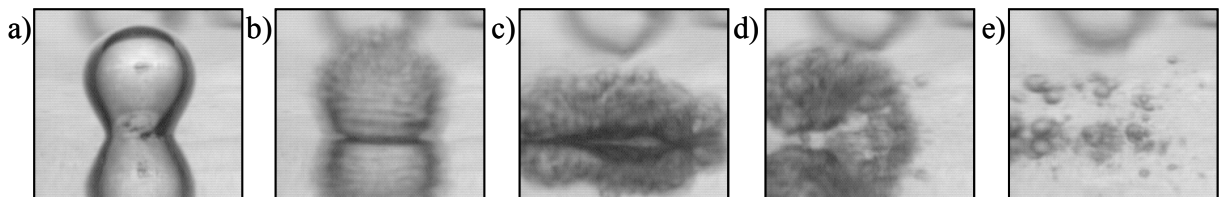

Fig. 5 Series of high-speed video frames showing a vapor bubble being removed from a boiling surface by an acoustic field. t/T = 0 (a), 0.4 (b), 1.2 (c), 2.7 (d), and 6.2 (e). (The bubble's reflection on the polished surface is visible).

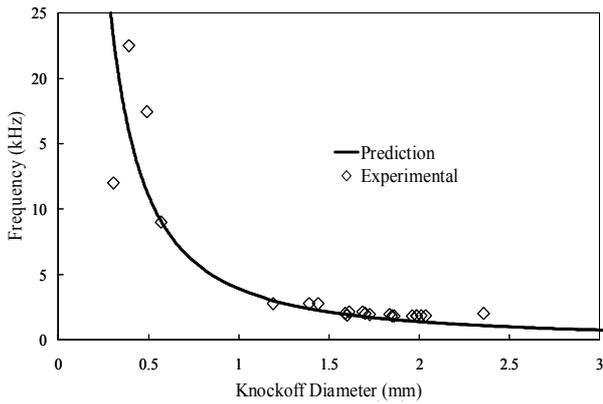

Fig. 6 Experimental observations (◊) and prediction (—) of knockoff diameter.

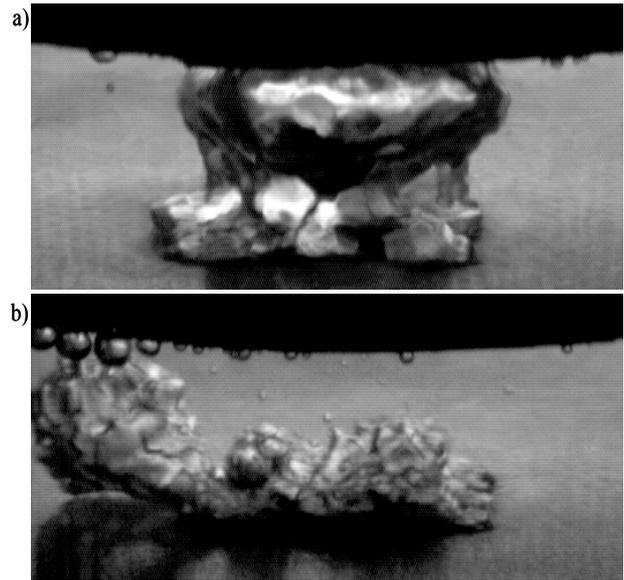

a)

b)

Fig. 8 Images of a) natural boiling immediately before CHF, and b) in the presence of a 1 kHz acoustic field at the same conditions.

For a majority of the experiments in this work the distance between the heated surface and the acoustic driver is 1 cm. One of the reasons for selecting this distance is that at closer distances the acoustic driver blocks the rising vapor from advecting away from the vicinity of the heated surface, resulting in the formation of a vapor bridge between the surface and the acoustic driver which insulates the surface thereby severely limiting the CHF. As reported earlier, when the separation distance is 1 cm, the unactuated CHF is 333 W/cm$^2$. This is only insignificantly different from the CHF measured when the acoustic driver is completely removed from the experiment (i.e., an infinite separation distance). By decreasing the separation distance to 5 mm, the unactuated heat flux is decreased to 227 W/cm$^2$. By further decreasing the separation distance to 3.5 mm, the unactuated CHF is only 137 W/cm$^2$.

At separation distances less than 1 cm, the improvement due to the acoustic enhancement is increased significantly in terms of percentage improvement. At a separation distance of 5 mm, the CHF with a 1 kHz, 70.5 kPa acoustic field is 368 W/cm$^2$. Although this is 36 W/cm$^2$ less than the CHF under the same conditions for a 1 cm separation distance, it is an increase of 62% over the 227 W/cm$^2$ measured without the acoustic field and a separation distance of 5 mm. At a separation distance of 3.5 mm, the increase is even more dramatic. The CHF with a 1 kHz, 70.5 kPa acoustic field is 338 W/cm$^2$, which is a 147% improvement over the same case without an acoustic field. The CHF and improvements at different separation distances are summarized in Table 1. These improvements could prove beneficial when space or other engineering requirements require a wall to be placed very near a heated surface cooled by two-phase boiling heat transfer.

Improvements in the CHF are realized for driving frequencies of 0.7 − 1.2 kHz (0.7 kHz is the lower limit for effective operation of the current acoustic driver). The exact reason why frequencies above 1.2 kHz do not have the same effect on the system as lower frequencies is not completely understood, but it is suspected that the sizes of the vapor bubbles and columns are larger than the size of the bubbles affected by the higher frequencies. The optimal frequency for removing a single bubble from the surface is given by (1). Thus, for 1.2 kHz, an "optimal" bubble size is approximately 2.2 mm. The size of the vapor structures in Fig. 8a is approximately 8 mm. Preliminary experiments, using a smaller 2.5 mm square heater, which produces smaller vapor structures, indicate that higher frequencies may be more effective on smaller scale processes.

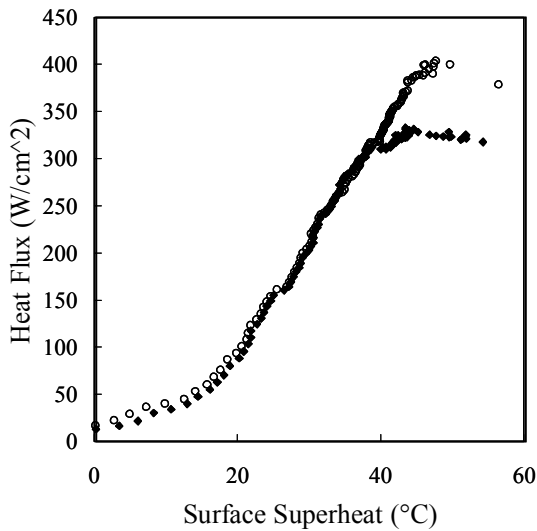

Fig. 7 Heat flux in the absence (♦) and presence (○) of 1 kHz acoustic field.

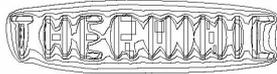

*Budapest, Hungary, 17-19 September 2007*

TABLE 1

COMPARISON OF NATURAL AND ACOUSTICALLY ENAHNCED CHFES AT DIFFERENT SEPARATION DISTANCES

| Separation Distance (mm) | CHF (W/cm$^2$) No Acoustic Field (Superheat) | CHF (W/cm$^2$) 1 kHz, 70.5 kPa Acoustic Field (Superheat) | Improvement |
|---|---|---|---|
| ∞ | 333 (40.1°C) | N/A | N/A |
| 10 | 333 (40.1°C) | 404 (49.7°C) | 21% |
| 5 | 227 (38.4°C) | 368 (28.7°C) | 62% |
| 3.5 | 165 (26.1°C) | 338 (32.4°C) | 147% |

## IV. CONCLUSIONS

The present work investigates the efficacy of using a small, light-weight, low-power acoustic driver to enhance boiling heat transfer from a flat surface by facilitating the removal of vapor bubbles from the surface and suppressing the instabilities that lead to the transition to film boiling at the CHF. The acoustic field produced by the driver induces interfacial instabilities that affect a bubble's contact line with the heated surface and result in bubble detachment. Once detached, the bubble is easily advected into the bulk liquid by acoustic or buoyant forces.

The effects of the acoustic field on single air and vapor bubbles are investigated in the presence of the acoustic field. Variables that affect the interfacial instabilities include the frequency and sound pressure amplitude of the acoustic field and the bubble diameter. Next, the effect of the acoustic field on the boiling process on a submerged heater is studied and it is found that acoustic actuation increases the CHF.

In the present setup, the CHF increases by 34% at 19.8°C subcooling. The acoustic field limits the formation of large vapor columns and delays transition from nucleate to film boiling. Acoustic enhancement is observed at frequencies between 0.7 and 1.2 kHz. The lower end of this frequency band is limited by the acoustic driver. Frequencies above 1.2 kHz are thought to be too high to adequately influence the vapor structures on a 1 cm$^2$ heater because these frequencies are outside the optimal frequency band for influencing the approximately 8 mm vapor structures that commonly occur. Preliminary experiments using a heater with a smaller exposed surface indicate that higher frequencies are effective on smaller heaters.

When the distance between the boiling surface and the 3.2 cm diameter driver is less than 1 cm, a vapor bridge forms between the two surfaces and results in a decrease of the CHF. At a separation distance of only 3.5 mm the unactuated CHF decreases to 165 W/cm$^2$. At the same separation distance, the acoustically enhanced CHF is 338 W/cm$^2$, which is a 147% improvement. Therefore, when a boiling surface is geometrically constrained, the acoustic enhancement can be beneficial in increasing the CHF.